# Evidence for Nodal superconductivity in $Sr_2ScFePO_3$

K A Yates, ITM Usman, K Morrison, JD Moore, A M Gilbertson, AD Caplin, LF Cohen
*Physics Department, The Blackett Laboratory, Imperial College London, UK, SW7 2AZ*

H. Ogino, J. Shimoyama
*Department of Applied Chemistry, The University of Tokyo, 7-3-1 Hongo, Bunkyo-ku, Tokyo 113-8656, Japan*

**Abstract**

Point contact Andreev reflection spectra have been taken as a function of temperature and magnetic field on the polycrystalline form of the newly discovered iron-based superconductor $Sr_2ScFePO_3$. A zero bias conductance peak which disappears at the superconducting transition temperature, dominates all of the spectra. Data taken in high magnetic fields show that this feature survives until 7T at 2K and a flattening of the feature is observed in some contacts. Here we inspect whether these observations can be interpreted within a d-wave, or nodal order parameter framework which would be consistent with the recent theoretical model where the height of the P in the Fe-P-Fe plane is key to the symmetry of the superconductivity. However, in polycrystalline samples care must be taken when examining Andreev spectra to eliminate or take into account artefacts associated with the possible effects of Josephson junctions and random alignment of grains.

The discovery last year of superconductivity at 26K in LaFeAsO$_{1-x}$F$_x$ [1] has provoked a surge of interest in the various pnictide materials both experimentally and theoretically. Very quickly, the parent compounds of the REFeAsO$_{1-x}$F$_x$ series (RE is a rare earth) were doped to produce a superconducting transition temperature of up to T$_c$ ~ 55K [2] and new families of pnictide material were discovered such as the "122" series eg. (Ba,K)(Fe,Co)$_2$As$_2$,[3]. Recently another family has been discovered, the highly layered "22426" family of which Sr$_2$ScFePO$_3$ has a T$_c$ of 17 K[4]. This family consists of FeP layers separated by Sr$_2$ScO$_3$ perovskite blocks and offers significant opportunities for doping both in terms of carrier density changes and structural effects [4]. Indeed, by substituting As for P and V for Sc, a T$_c$ of 37K has been reported [5]. Although significant experimental progress has been made in the past year, there is still wide theoretical discussion about the superconducting order parameter symmetry, the number of gaps involved in the superconductivity and the role of spin fluctuations in mediating the superconductivity, for a review see [6].

Although the theories vary in the details, the complex Fermi surfaces of the pnictide materials are known to play a key role in the superconductivity. For LaFePO the Fermi surface is composed of two hole like surfaces around the Γ point and two electron like Fermi surfaces around the M point [6, 7]. The Fermi surface of Sr$_2$ScFePO$_3$ is predicted to be similar but with the significant difference that one of the electron bands is significantly smaller than the other [8] and so the FS nesting is significantly reduced in this material. When optimally doped, the majority of theories predict that fully gapped s-wave order parameters (OP) open on both the hole and electron Fermi sheets below T$_c$ but that they are π phase shifted with respect to each other. This model has been termed the "extended s-wave" or "s±" state [6]. It has been shown that the fully gapped s± state is in competition with a d-wave or nodal s± OP [9,10] and, it has been suggested that the pnictogen height in the Fe-Pn-Fe plane will act as a switch between nodeless s± and nodal superconductivity [10]. Experimentally it has been shown that the angle of the Fe-Pn-Fe bond also plays a key role in the $T_c$ of the material [11]. In this context the $Sr_2ScFePO_3$ compound is of particular interest. The combination of the angle of the Fe-Pn-Fe bond (118°) and the a-axis lattice parameter (4.016Å) means that the height of the P above the Fe plane is 1.20Å [12] this is lower than the As height in LaFeAsOF (1.32 Å) and indeed more comparable, although still higher than, the P height in LaFePO (1.12 Å), a material recently shown to have a nodal OP [13,14]. The material therefore is a good test of whether the pnictogen height may play a role in the superconducting symmetry. Point contact Andreev reflection (PCAR) is a sensitive probe of the OP symmetry and the number of gaps in novel superconductors [15,16]. Although for the pnictide superconductors, the results of PCAR have been contradictory [17,18,19,20,21,22,23,24] the majority of studies on the optimally doped materials favour a nodeless OP [18,22,23,24] although some zero bias conductance peaks (zbcp) in the spectra have been observed, which could indicate a d-wave OP [20,24]. To a certain extent, the differences in the spectra may be attributable to the multiband properties of the pnictides and indeed may even be expected if the superconductor has s± symmetry [25,26,27]. Nonetheless a true d-wave OP produces conductance spectra characterised by more than just the zbcp. In particular, the magnetic field dependence of the zbcp can be a key indicator of the OP symmetry [15]. In this *paper* we show point contact spectra as a function of temperature and magnetic field and show that the newly discovered $Sr_2ScFePO_3$ material is most probably a nodal superconductor.

The polycrystalline materials studied here were prepared by solid state reaction as described in ref. [4]. X-ray diffraction indicated a small amount of secondary phase identified as $SrFe_2P_2$. The transition temperature of the sample studied was determined resistively and by magnetization measurements to be 15 K. Scanning electron microscope (SEM) images indicated randomly orientated grains with a grain size of 2-20 µm. Point contact spectra were obtained by using either 'hard' or 'soft' geometry. For the hard point contact (HPC), a mechanically sharpened Au tip was brought into contact with the sample and the conductance was measured differentially across the tip-sample contact, as described in [16]. For the soft point contact (SPC), the method of ref [22] was used whereby a drop of silver paint formed the contact and the differential resistance was measured across that contact. Due to the nanocrystalline nature of the silver paint, the contact is formed by many nanocontacts analogously to the HPC technique and has been shown to be very successful for both $MgB_2$ [28] and the oxypnictide materials [22,23]. Both HPC and SPC measurements were performed as a function of temperature and magnetic field. Spectra were fitted to either the Blonder Tinkham Klapwijk (BTK) model [29] which assumes a single s-wave gap, to the BTK model adapted for two gaps as used for $MgB_2$ [16] or to the the Kashiwaya-Tanaka adaptation of the BTK model which is applicable to nodal superconductors such as $d_{x2-y2}$ order parameters [30]. In addition to the gap value, $\Delta$, the interface scattering, Z and the smearing parameter, $\omega$, which includes both thermal and non-thermal contributions, the Kashiwaya-Tanaka model introduces an effective in-plane angle at which the point contact current enters the nodal superconductor, $\alpha$.

One consequence of a d-wave or nodal-with sign change OP is that a zbcp appears in the PCAR conductance spectra [30]. The effect of the angle of incidence of the point contact current assuming a $d_{x2-y2}$ OP is shown for two contacts of different Z in the generated spectra in figure 1. Given the random orientation and small size of the grains observed in the SEM and that the average 'foot print' of a HPC is 20-50µm although this consists of many individual nano-contacts [16], it is likely if the superconductivity has this symmetry that the zbcp would dominate any spectrum that effectively averages over the contributions across all of the grains. The result of such a simple average, where the conductance contributions are added with equal weight across all $\alpha$ is shown in the inset to figure 1.

The experimentally determined temperature dependence of a typical HPC spectrum (contact HPC3) is shown in figure 2a. The low temperature data show a pronounced zbcp with shoulders at ~ 7mV similar to that observed in the generated spectra in figure 1a. As the temperature increases, the height of the zbcp decreases until, at a temperature of 15.3K the conductance spectrum is flat, indicating the $T_c$ for that contact. It is important to eliminate thermal effects where possible. It is well appreciated that both hard and soft point contacts are made up of many parallel channels of nano-contact dimension [16,22] much smaller than the overall footprint of the contact. Using the Sharvin or Wexler formulae [33], estimates can be made of the effective contact size. If we use the resistivity $\rho$ and mean free path $l$ of Au (which are 2.2 x$10^{-10}$ $\Omega$m and 3.84x$10^{-6}$ m at 4.2K respectively), the effective contact dimension would appear to be between 3 - 9 nm. If a weighted average of the resistivity of the two materials is used [35] the estimated contact size will be larger, of the order of 90 - 300nm. If the formulae are used with a weighted average of the mean free paths of both materials, the effective contact size approaches 1 micron. The mean free path reported for 122 pnictides is $l \sim$ 8nm [34]), and comparing this directly with the effective contact size would suggest that the measurements are close to or in the diffusive regime. However, it is well known that this effective contact 'size' is composed of many nano-contacts in parallel [16,22] and it is therefore not possible to say whether the individual contacts are in the ballistic or diffusive regime [22]. Nevertheless within the general formulation of Andreev reflection for a single s-wave superconductor [36], a zero bias conductance peak is not anticipated to occur under the conditions of ballistic or diffusive contacts. Diffusive contacts suppress the Andreev doubling of the conductance for bias voltages less than the gap voltage but do not lead to zbcp features within any usual range of interface parameters. We can state definitively however that most contacts are clearly not in the thermal limit. The existence of the zbcp until T~Tc and the coincidence of the conductance spectra at high bias voltage (inset to fig. 2a) indicates that contact heating is not a significant problem for the results presented here. Note that the background conductance of SPC2 may imply moderate heating may be occurring in that contact, and we include that data in order to show that these types of spectra occur but not in the majority of contacts formed. SPC spectra are shown in figures 2b (contact SPC1) and 2c (contact SPC2). Again a prominent zbcp is observed which decreases in intensity until at ~16K the spectrum is that of the background, indicating a $T_c$ between 15 and 17K in those contacts. It has been suggested that instead of being attributable to a d-wave OP, zbcp in HPC spectra onto polycrystalline materials can be artefact related to the mechanical nature of the contact and the fact that it pushes into the granular sample. The observation of prominent zbcp in both HPC and SPC spectra strongly suggest that the zbcp observed here are intrinsic to the sample rather than an artefact associated with the granularity. The spectra shown in figures 2, 4 and 5 represent only a fraction of the total contacts made to two samples (A and B) of the same composition. To illustrate the range of spectra obtained, a variety of different contacts are shown in figure 3 at low temperature.

We show our attempt to fit the data using the d-wave model of Kashiwaya-Tanaka in figure 4. To avoid the potential degeneracy of a four parameter fit, the data were fitted with the Kashiwaya-Tanaka model [30] and with $\Delta$, $Z$, $\omega$, to give the least squares fit, $\chi^2$, for fixed but incrementally increasing, values of $\alpha$. A plot of $\chi^2(\alpha)$ then gave the best fit as detailed in ref [24]. It can be seen that the fitting procedure is not able to follow the HPC shape of the zbcp with shoulders perfectly, (figure 4a) although a value for the gap parameter ($\Delta = 4.34 \pm 0.04$meV, $2\Delta/kT_c = 6.7$) consistent with reports on other pnictides can be obtained. Forcing the fit to the zbcp (by reducing the range over which $\alpha$ was varied, dotted line in fig 4a) does improve the quality of the fit, however, the gap value is now overestimated ($\Delta = 7.86 \pm 0.07$meV, $2\Delta/kT_c = 11$). The SPC1 (SPC2) contacts shown in figure 2b,(c) could be fitted if $\Delta = 7.01,(1.36)$ meV, $Z = 1.28,(0.88)$ $\omega = 2.98$ (2.08) meV and $\alpha = 0.435$ (310) rad. Clearly our attempts to fit a d-wave model is unsatisfactory.

S-wave and multiple s-wave models do not satisfactorily account for the zbcp. In fact the data look remarkably similar to data taken on HTS cuprate polycrystalline material [31 –see fig 7 in that reference], which were interpreted in terms of a d wave model. Although the zbcp is expected in some models for Andreev into the s± state, it is only expected to occur when certain conditions are met regarding the tunnelling probabilities into each of the bands [25]. The fact that the zbcp is consistently observed in the spectra taken on the $Sr_2ScFePO_3$ indicate that this a less attractive explanation for the spectra presented here.

It is important to eliminate weak links as a potential cause for the zbcp observed in these spectra. In order to address this issue, the magnetic field dependence is shown in figure 5 for HPC (contact HPC5) (fig 5a) and SPC (SPC1,2) (fig 5b,c) contacts. There is an initial sharp decrease in zbcp in low field as shown explicitly in figure 6, which could be associated with weak links, however the feature survives up to high fields, suggesting that inter-granular or intrinsic (ie due to the natural layering of the compound) Josephson junctions [32], cannot be the sole explanation. The field at which the zbcp disappears in the current spectra is coincident with the field at which superconductivity is destroyed in the sample i.e. $H_{c2}$ [4]. Furthermore, in the HPC and the SPC1, a flattening of the zbcp is observed with increasing field. Such a flattening may be expected if the zbcp was being split by the magnetic field (as has been observed previously in thin films of the cuprate superconductor YBCO [15]) but the contact here is averaged over many grains of different orientation, as suggested by the SEM image of the sample (Andreev footprint is

50 μm, grain size of sample is on average ~2 μm). Not all contacts showed such significant flattening, perhaps consistent with the fact that different contacts will average over different orientations of the grains. In order to satisfactorily explain whether this flattening in field is a definitive suggestion of a nodal OP with sign-change, further work on single crystals of this material will be needed.

In conclusion, we have studied the newly discovered superconductor $Sr_2ScFePO_3$ and find that the data is shows spectra features indicative of a superconductor with a nodal order parameter. Measurements of the spectra as a function of magnetic field show that this zbcp feature survives up to high magnetic field suggesting that Josephson effects are unlikely to be the dominant explanation. Heating effects can also be ruled out as the low temperature conductance curves merge with the above $T_c$ curves at high voltage bias. The spectra cannot be fitted assuming a singular or multiple s-wave model while the s± model is unlikely to hold due to the ubiquity of the zbcp. The presence of a nodal order parameter in this material is consistent with the proposed model of the pnictogen height acting as a switch between nodal and nodeless OPs [10] and suggests that further work on single crystals in which the pnictogen height is systematically varied would be extremely interesting.

**Acknowledgements**: We would like to thank Y. Matsumura for sample preparation and RC Maher, Werner J Karl, Anisha Mukherjee for SEM measurements.

**Figure captions**

Figure 1: Effect of the angle of incidence, $\alpha$, on the generated conductance spectra for $\alpha = 0.07$ (black solid line) to $\alpha = 0.57$ (dotted line) in steps of $\alpha = 0.1$rad. Other parameters used to model the data are $\Delta = 5$meV, $\omega = 0.53$meV and (a) $Z = 0.4$, (b) $Z = 1.0$. Inset shows the results of a simple average of conductance spectra for both values of Z and $\alpha = 0.07$-1.57rad

Figure 2: Temperature dependence of (a) an HPC (HPC3) at 4.2K (top spectrum), 5.2K, 6.9K, 8.1K, 9.5K, 11.2K, 12.5K and 15.3K (bottom spectrum) (b) an SPC (SPC1) at 2 K (top spectrum), 3K, 5K, 7K, 9K, 11K, 13K and 15K (bottom spectrum). (c) an SPC (SPC2) at 2K (bottom spectrum), 4K, 6K, 8K, 10K, 12K, 14K, 15K, 16K and 17K. The data are not offset. Inset to figure (a) shows conductance curve of the HPC at 4.2K and 15.3K indicating that the conductance enhancement occurs over a wide voltage range. Due to the limited range for the 15.3K spectrum the dotted line indicates the extrapolation of the data.

Figure 3: Typical spectra from a variety of different HPC (at 4.2K) and SPC (at 4K) contacts. (a) HPC contact 1 (sample A), (b) HPC 3 (sample A), (c) HPC 2 (sample A), (d) HPC 4 (sample B) (e) SPC1 (sample A), (f) SPC 2 (sample A).

Figure 4: Results of fitting (a) the 4.2K HPC spectrum in figure 2 (HPC 3). Allowing all parameters to vary (solid line) $\Delta = 4.34 \pm 0.04$meV, $Z = 0.45 \pm 0.02$, $\omega = 2.29$ meV, $\alpha = 0.405 \pm 0.005$rad; limiting $\alpha$ to force a better fit (dashed line) $\Delta = 7.86 \pm 0.07$ meV, $Z =$

0.73 ± 0.01, $\omega$ = 1.62 ± 0.01meV, $\alpha$ = 0.382 ± 0.002rad. (b) The 2K spectrum (SPC1) with $\Delta$ = 1.36 meV, Z = 0.88 meV, $\omega$ = 2.08 meV, $\alpha$ = 0.435 rad. (c) The 2K SPC spectrum (SPC2), with $\Delta$ = 7.01 meV, Z = 1.28 meV, $\omega$ = 2.98 meV, $\alpha$ = 0.435 rad.

Figure 5: Magnetic field dependence of (a) HPC (HPC5) at 9K normalised to 20mV at 0, 1, 2, 3, 8T T (top to bottom), dashed line is a fit to the 0T spectrum with assuming a $d_{x2-y2}$ order parameter and $\alpha$ = 0.364 rad, $\Delta$ = 2.97,Z = 0.72,$\omega$ = 2.16meV, (b) SPC (SPC1) at 4K normalised to 20mV at 0,1,2,3,4,5,7 T (top to bottom), dashed line is a fit to the 0T spectrum with $\alpha$ = 0.344 rad, $\Delta$ = 1.61,Z = 0.92,$\omega$ = 2.27meV (c) SPC (SPC2) at 4K normalised to 15mV at 0, 0.4, 1.2, 2.0, 2.8, 3.6, 4.4, 5.2, 6.0, 6.8T (top to bottom). Dotted line shows fit with $\alpha$ = 0.435 rad, $\Delta$ = 6.61 meV, Z = 1.28, $\omega$ = 3.24meV. Inset shows HPC5 spectra at 0 and 8T over an extended bias range.

Figure 6: Conductance at zero bias as a function of magnetic field for (a) HPC5 at 9K, (b) SPC1 (○) and SPC2 (■) at 4K.

[1] Y Kamihara et al, J Am Ceram Soc, 130, 3296 (2008)
[2] Ren ZA, Lu W, Yang J, Yi W, Shen XL, Li ZC, et al. Chinese Physics Letters;**25** (6):2215-6. (2008)

[3] M Rotter, M Tegel, D Johrendt et al, Phys Rev. B, **78**, 020503 (2008)
[4] H Ogino, Y Matsumura, Y Katsura et al, Supercond. Sci. Technol, 22, 075008 (2009)
[5] X Zhu, F Han, G. Mu et al, Phys. Rev. B, 79, 220512(R) (2009)
[6] I.I. Mazin, J. Schmalian, Physica C, **469**, 614 (2009)
[7] A.I. Coldea, J.D. Fletcher, A. Carrington et al, Phys. Rev. Lett, **101**, 216402 (2008)
[8] I.R. Shein, A.L. Ivanovskii, Phys Rev B, **79**, 245115 (2009)
[9] S Graser, T.A. Maier, P.J. Hirschfeld, D.J. Scalapino, New J Phys, 11, 025016 (2009)
[10] K.Kuroki, H. Usui, S Onari, R. Arita, H. Aoki, Phys. Rev. B, 79, 224511 (2009)
[11] C-H Lee, A. Iyo, H. Eisaki et al, J. Phys. Soc. Jpn, 77, 083704 (2008)
[12] Kotegawa et al, arxiv : 0908.1469 (2009)
[13] J.D. Fletcher, A. Serafin, L. Malone et al, Phys. Rev. Lett, 102, 147001 (2009)
[14] CW Hicks et al,Phys Rev Lett, **103**, 127003 (2009)
[15] G Deutscher, Rev. Mod. Phys, 77, 109 (2005)
[16] Y. Bugoslavsky, Y. Miyoshi, G.K. Perkins et al, Phys. Rev. B, 72, 224506 (2005)
[17] Y.L. Wang, L. Shan, L. Fang et al, Supercond Sci Technol, 22, 015018 (2009)
[18] P Szabo, Z Pribulova, G. Pristas, et al, Phys. Rev. B, 79, 012503 (2009)
[19] Y.T. Chen, Z. Tersanovic, R.H. Liu et al, Nature, 453, 1224 (2008)
[20] L Shan, Y Wang, Z Zhu et al, Europhys Lett, 83, 57004 (2008)
[21] P Samuely, P Szabo, Z Pribulova, Supercond Sci Technol, 22, 014003 (2009)
[22] R.S. Gonnelli, D. Daghero, M. Tortello et al, Phys. Rev. B, 79, 184526 (2009)
[23] D Daghero, M Tortello, R.S. Gonnelli et al, Phys. Rev. B, 80, 060502 (2009)
[24] K.A. Yates et al, New J Phys, 11, 025015 (2009)
[25] A A Golubov, A Brinkman, OV. Dolgov et al, Phys Rev Lett, 103, 077003 (2009)
[26] P Ghaemi, F Wang, A Vishwanath, Phys Rev Lett, 102, 157002 (2009)
[27] M.A.N Araujo, P.D. Sacramento, Phys Rev B 79, 174529 (2009)
[28] R.S. Gonnelli, D Daghero, A. Calzolari et al, Phys Rev. B 71, 060503(R) (2005)
[29] G. E. Blonder, M. Tinkham, and T. M. Klapwijk, Phys Rev B **25**, 4515 (1982).
[30] S Kashiwaya, Y Tanaka, M Koysnagi et al, Phys Rev B, 51, 1350 (1995)
[31] Dagan and Deutscher, Phys Rev B, 61, 7015 (2000)
[32] Kleiner, J. Low Temp Phys, 106, 314 (1997)
[33] Duif et al, J Phys Cond Mat. 1, 3157, (1989)
[34] Yin et al, Phys Rev Lett, 102, 097002 (2009)
[35] Sheet et al, Phys Rev B, 69, 134507 (2004)
[36] I.I. Mazin, A.A. Golubov, B. Nadgorny, J Appl Phys, 89, 7576, (2001)

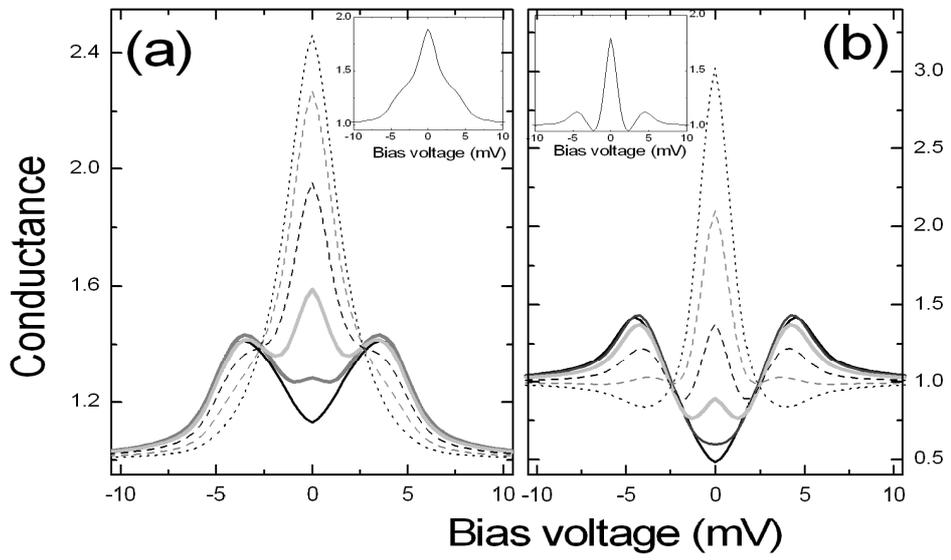

Figure 1

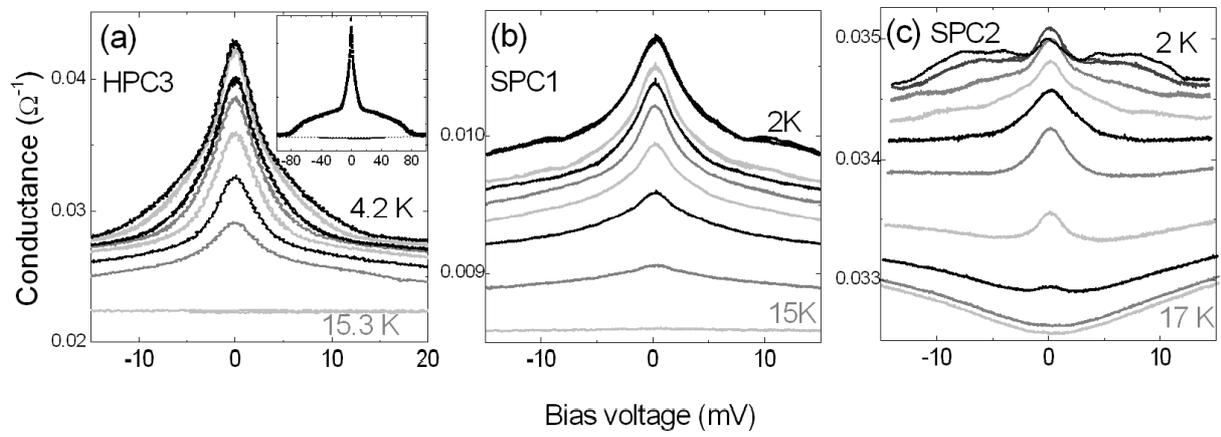

Figure 2

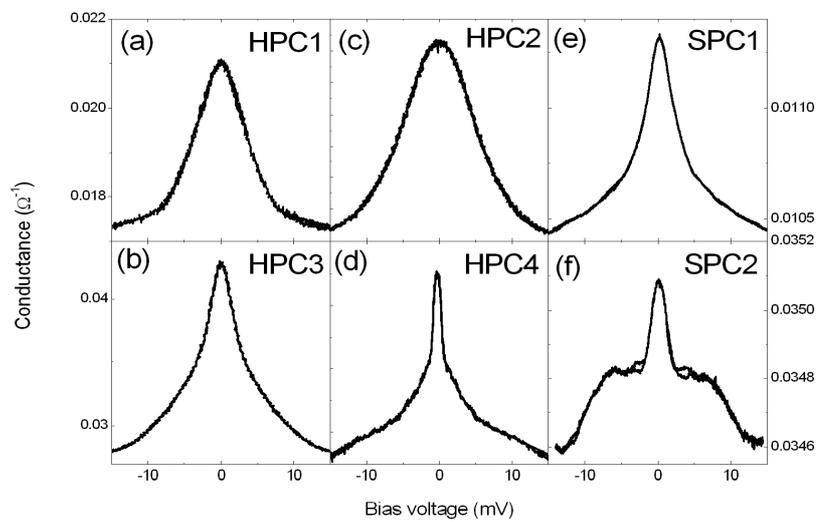

Figure 3

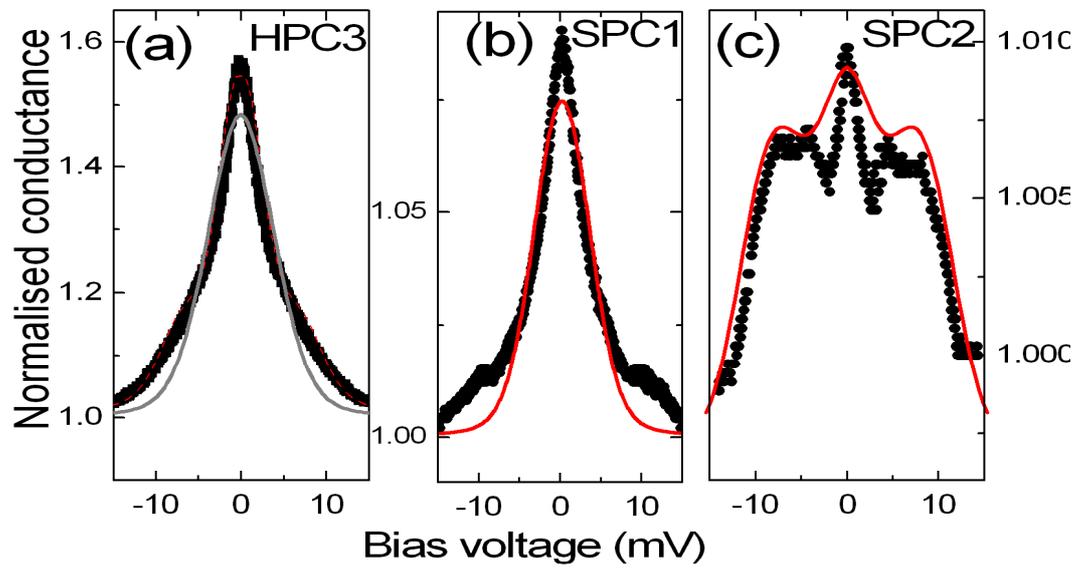

Figure 4

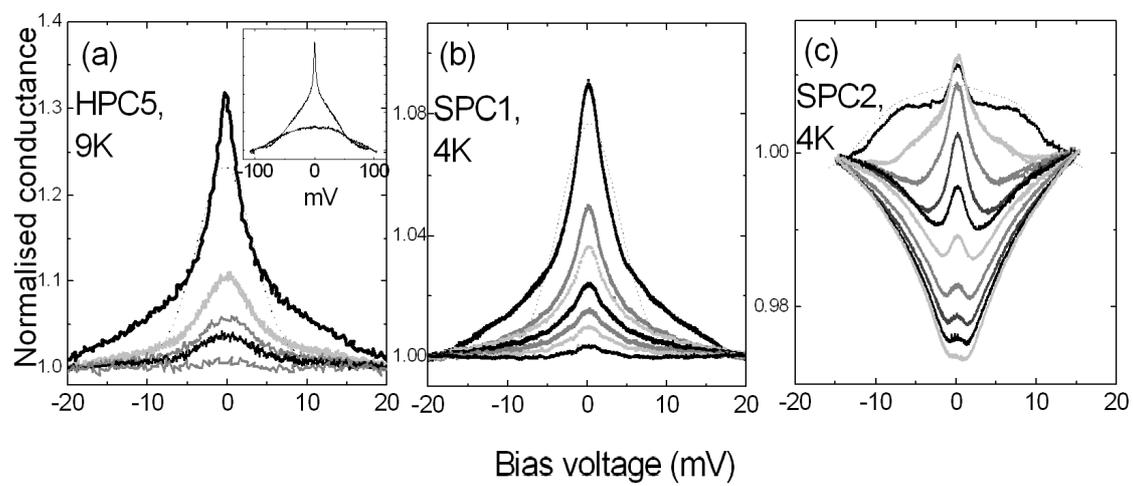

Figure 5

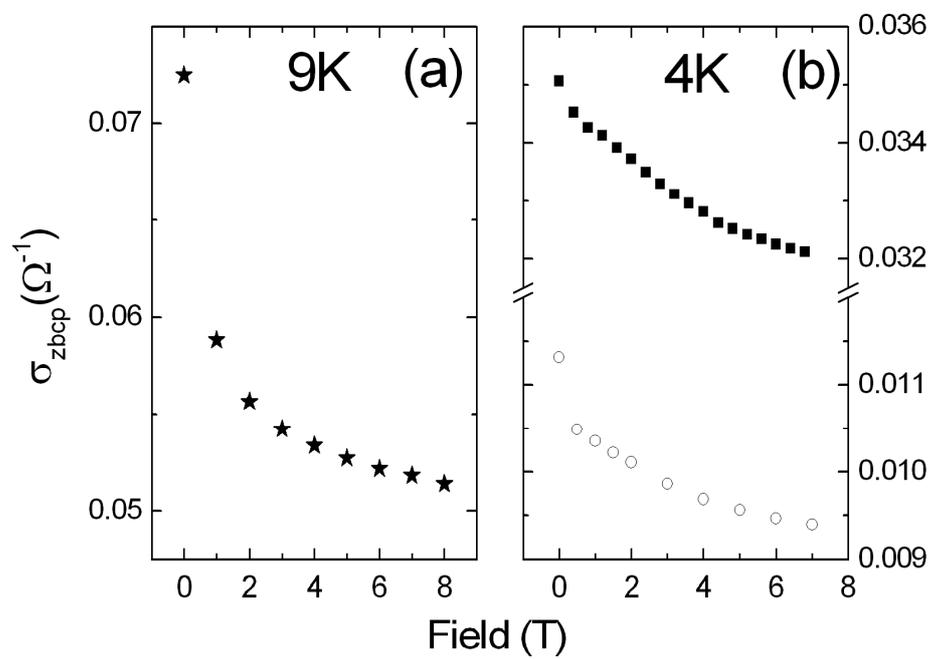

Figure 6